# Axioms and uniqueness theorem for Tsallis entropy

Sumiyoshi Abe

*College of Science and Technology, Nihon University,*
*Funabashi, Chiba 274-8501, Japan*

The Shannon-Khinchin axioms for the ordinary information entropy are generalized in a natural way to the nonextensive systems based on the concept of nonextensive conditional entropy and a complete proof of the uniqueness theorem for the Tsallis entropy is presented. This improves the discussion of dos Santos.

# 1. Introduction

The concept of entropy is the fundamental ingredient in information theory and statistical mechanics. Given a probability distribution $\{p_i\}_{i=1,2,\ldots,W}$ satisfying $0 \leq p_i$ $(i=1,2,\ldots,W)$ and $\sum_{i=1}^{W} p_i = 1$, the Boltzmann-Shannon entropy, i.e., the ordinary information entropy, is defined by

$$S(p_1, p_2, \ldots, p_W) = -k \sum_{i=1}^{W} p_i \ln p_i, \qquad (1)$$

where $k$ is a positive constant, which is henceforth set equal to unity for simplicity. This quantity is a positive and concave function of $\{p_i\}_{i=1,2,\ldots,W}$. It also fulfills the H-theorem. Its maximum is realized by the equiprobability distribution, i.e., $p_i = 1/W$ $(i=1,2,\ldots,W)$ and the value itself is $S = \ln W$, which is the celebrated Boltzmann formula.

To characterize what the entropy is, Shannon [1] and Khinchin [2] investigated its axiomatic foundation. The Shannon-Khinchin axioms are given as follows:

[ ] $S(p_1, p_2, \ldots, p_W)$ is continuous with respect to all its arguments and takes its maximum for the equiprobability distribution $p_i = 1/W$ $(i=1,2,\ldots,W)$,

[ ] $S[A, B] = S[A] + S[B|A]$,

[ i] $S(p_1, p_2, \ldots, p_W, 0) = S(p_1, p_2, \ldots, p_W)$.

In the second axiom, $S[A, B]$ and $S[A]$ are the entropies of the composite system

$(A, B)$ with the joint probability distribution $p_{ij}(A, B)$ $(i = 1, 2, \ldots, W; j = 1, 2, \ldots, U)$ and the subsystem $A$ with the marginal probability distribution $p_i(A) = \sum_{j=1}^{U} p_{ij}(A, B)$, respectively. $S[B|A]$ stands for the conditional entropy associated with the conditional probability distribution $p_{ij}(B|A) = p_{ij}(A, B)/p_i(A)$. Writing the entropy of $p_{ij}(B|A)$ as $S[B|A_i]$, the conditional entropy is given by

$$S[B|A] = \langle S[B|A_i] \rangle^{(A)} = \sum_{i=1}^{W} p_i(A) S[B|A_i]. \qquad (2)$$

In the special case when $A$ and $B$ are statistically independent, $S[B|A] = S[B]$, leading to the additivity:

$$S[A, B] = S[A] + S[B]. \qquad (3)$$

The uniqueness theorem [1,2] states that a quantity satisfying the axioms [ ]-[ ¡] is uniquely equal to $S$ in eq. (1).

We note that there is a correspondence relation between the Bayes multiplication law and the axiom [ ]:

$$p_{ij}(A, B) = p_i(A) p_{ij}(B|A) \quad \leftrightarrow \quad S[A, B] = S[A] + S[B|A]. \qquad (4)$$

Now, in the field of statistical mechanics, there is a growing interest in nonextensive generalization of Boltzmann-Gibbs theory. In particular, the one initiated by Tsallis [3-5] is receiving much attention. This formalism is based on the following single-parameter generalization of the Boltzmann-Shannon entropy:

$$S_q(p_1, p_2, \ldots, p_W) = \frac{1}{1-q}\left[\sum_{i=1}^{W} (p_i)^q - 1\right] \quad (q > 0). \tag{5}$$

In the limit $q \to 1$, $S_q$ converges to $S$ in eq. (1) (with $k \equiv 1$). This quantity is also positive and concave, and satisfies the H-theorem. In addition, as the Boltzmann-Shannon entropy, it takes its maximum for the equiprobability distribution $p_i = 1/W$ ($i = 1, 2, \ldots, W$): Its value is $S_q = (1-q)^{-1}(W^{1-q} - 1)$, which is a monotonically increasing function of $W$. However, the additivity is violated. For statistically independent systems, $A$ and $B$, $S_q$ yields

$$S_q[A, B] = S_q[A] + S_q[B] + (1-q)S_q[A]S_q[B], \tag{6}$$

which is referred to as the pseudoadditivity. Clearly, the additivity holds only in the limit $q \to 1$. Equation (6) has been carefully discussed in Ref. [6].

In a recent paper [7], dos Santos has made an interesting discussion about uniqueness of the Tsallis entropy in eq. (5). He has shown that that if a quantity is [ ⅰ] continuous with respect to the probability distribution $\{p_i\}_{i=1,2,\ldots,W}$, [ ⅱ] a monotonically increasing function of $W$ in the case of the equiprobability distribution $p_i = 1/W$ ($i = 1, 2, \ldots, W$), and [ ⅲ] satisfies the pseudoadditivity in eq. (6) for the statistically independent systems, $A$ and $B$, and [ ¶ ] fulfills the relation $S_q(p_1, p_2, \ldots, p_W) = S_q(p_L, p_M) + (p_L)^q S_q(p_1/p_L, p_2/p_L, \ldots, p_W/p_L) + (p_M)^q S_q(p_1/p_M, p_2/p_M, \ldots, p_W/p_M)$, where $p_L = \sum_{i=1}^{W_L} p_i$ and $p_M = \sum_{i=W_L+1}^{W} p_i$, then it is identical to the Tsallis entropy.

Here, comparing the set of the Shannon-Khinchin axioms [ ]-[ ¡] with that of [ ╫]-[ ¶], we find that the complete parallelism is missing between the two. The main reason for this is due to the absence of the concept of "nonextensive conditional entropy".

In this paper, we present the axioms for the Tsallis entropy by introducing the nonextensive conditional entropy. Then we prove the uniqueness theorem for the Tsallis entropy. Our proof essentially follows a natural generalization of the line presented in Ref. [2]. This improves the discussion given in Ref. [7] and establishes the complete parallelism with the Shannon-Khinchin axioms.

## 2. Nonextensive conditional entropy

In nonextensive statistical mechanics, it is known [4] that the average of a physical quantity $Q = \{Q_i\}_{i=1,2,\text{L},W}$ is given in terms of the normalized $q$-expectation value:

$$\langle Q \rangle_q = \sum_{i=1}^{W} Q_i P_i \equiv \frac{\sum_{i=1}^{W} Q_i (p_i)^q}{\sum_{i=1}^{W} (p_i)^q}, \tag{7}$$

where $P_i \equiv (p_i)^q / \sum_{i=1}^{W} (p_i)^q$ is the escort distribution associated with $p_i$ [8]. To be consistent with the nonextensive formalism, we use this concept to generalize the definition in eq. (2). For this purpose, we calculate the Tsallis entropy of the conditional probability distribution

$$S_q[B|A_i] = \frac{1}{1-q}\left\{\sum_{j=1}^{U}\left[p_{ij}(B|A)\right]^q - 1\right\}. \tag{8}$$

From this quantity, we define the nonextensive conditional entropy as follows:

$$S_q[B|A] = \left\langle S_q[B|A_i]\right\rangle_q^{(A)} = \frac{\sum_{i=1}^{W}[p_i(A)]^q S_q[B|A_i]}{\sum_{i=1}^{W}[p_i(A)]^q}. \tag{9}$$

Using the definition of the Tsallis entropy, we find that equation (9) can be expressed as

$$S_q[B|A] = \frac{S_q[A, B] - S_q[A]}{1 + (1-q)S_q[A]}. \tag{10}$$

An important point here is that, with this definition, a natural nonextensive generalization of the correspondence relation in eq. (4) is established in conformity with the pseudoadditivity in eq. (6):

$$p_{ij}(A, B) = p_i(A) p_{ij}(B|A)$$
$$\leftrightarrow S_q[A, B] = S_q[A] + S_q[B|A] + (1-q)S_q[A]S_q[B|A]. \tag{11}$$

In fact, the pseudoadditivity in eq. (6) is recovered in the special case when $A$ and $B$ are statistically independent each other.

## 3. Axioms and uniqueness theorem for Tsallis entropy

Now, the set of the axioms we present for the Tsallis entropy is the following:

[ ]*   $S_q(p_1, p_2, \ldots, p_W)$ is continuous with respect to all its arguments
and takes its maximum for the equiprobability distribution $p_i = 1/W$
$(i = 1, 2, \ldots, W)$,

[ ]*   $S_q[A, B] = S_q[A] + S_q[B|A] + (1-q) S_q[A] S_q[B|A]$,

[ ¡]*   $S_q(p_1, p_2, \ldots, p_W, 0) = S_q(p_1, p_2, \ldots, p_W)$.

**Theorem:** A quantity satisfying [ ]*-[ ¡]* is uniquely equal to the Tsallis entropy.

*Proof:* First, let us consider the equiprobability distribution $p_i = 1/W$ $(i = 1, 2, \ldots, W)$ and put

$$S_q\left(\frac{1}{W}, \frac{1}{W}, \ldots, \frac{1}{W}\right) := L_q(W). \qquad (12)$$

From [ ¡]*, it follows that

$$\begin{aligned} L_q(W) &= S_q\left(\frac{1}{W}, \frac{1}{W}, \ldots, \frac{1}{W}, 0\right) \\ &\leq S_q\left(\frac{1}{W+1}, \frac{1}{W+1}, \ldots, \frac{1}{W+1}, \frac{1}{W+1}\right) = L_q(W+1), \end{aligned} \qquad (13)$$

which means that $L_q(W)$ is a nondecreasing function of $W$.

Consider $m$ statistically independent systems, $A_1, A_2, \ldots, A_m$, each of which contains $r \, (\geq 2)$ equally likely events. Then, we have

$$S_q[A_k] = S_q\left(\frac{1}{r}, \frac{1}{r}, \ldots, \frac{1}{r}\right) = L_q(r) \qquad (1 \leq k \leq m). \tag{14}$$

Using [ ]* for these independent systems, we find

$$S_q[A_1, A_2, \ldots, A_m] = \sum_{k=1}^{m} \binom{m}{k} (1-q)^{k-1} [L_q(r)]^k$$

$$= \frac{1}{1-q}\left\{[1 + (1-q) L_q(r)]^m - 1\right\}. \tag{15}$$

Since $S_q[A_1, A_2, \ldots, A_m] = L_q(r^m)$, we have

$$L_q(r^m) = \frac{1}{1-q}\left\{[1 + (1-q) L_q(r)]^m - 1\right\}. \tag{16}$$

Similarly, for any other positive integers $n$ and $s$ equal to or larger than 2, we have

$$L_q(s^n) = \frac{1}{1-q}\left\{[1 + (1-q) L_q(s)]^n - 1\right\}. \tag{17}$$

It is always possible to take $m$, $r$, $n$, and $s$ which satisfy

$$r^m \leq s^n \leq r^{m+1}. \tag{18}$$

Since $L_q$ is a nondecreasing function, we have the inequalities

$$L_q(r^m) \leq L_q(s^n) \leq L_q(r^{m+1}), \tag{19}$$

which lead to

$$\frac{1}{1-q}\left\{[1+(1-q)L_q(r)]^m - 1\right\} \leq \frac{1}{1-q}\left\{[1+(1-q)L_q(s)]^n - 1\right\}$$

$$\leq \frac{1}{1-q}\left\{[1+(1-q)L_q(r)]^{m+1} - 1\right\}. \tag{20}$$

Here, it is necessary to examine two cases, $0 < q < 1$ and $q > 1$, separately. After simple algebra, we find that in both cases the following inequalities hold:

$$\frac{m}{n} \leq \frac{\ln[1+(1-q)L_q(s)]}{\ln[1+(1-q)L_q(r)]} \leq \frac{m}{n} + \frac{1}{n}. \tag{21}$$

We note that, in deriving these inequalities, the following Ansatz has to be made:

$$0 < 1+(1-q)L_q(r), \ 1+(1-q)L_q(s) < 1 \quad \text{for } q > 1. \tag{22}$$

Later, we shall see that this is in fact justified. Also, from eq. (18), it is evident that

$$\frac{m}{n} \leq \frac{\ln s}{\ln r} \leq \frac{m}{n} + \frac{1}{n}. \tag{23}$$

Combining this with eq. (21), we have

$$\left| \frac{\ln\left[1+(1-q)L_q(s)\right]}{\ln\left[1+(1-q)L_q(r)\right]} - \frac{\ln s}{\ln r} \right| \leq \frac{1}{n}. \tag{24}$$

Since $n$ can be arbitrarily large, we obtain

$$\frac{\ln\left[1+(1-q)L_q(r)\right]}{\ln r} = \frac{\ln\left[1+(1-q)L_q(s)\right]}{\ln s} := \lambda(q), \tag{25}$$

where $\lambda(q)$ is a separation constant dependent on $q$. Therefore, we find

$$L_q(r) = \frac{1}{1-q}\left[r^{\lambda(q)} - 1\right]. \tag{26}$$

Clearly, $\lambda(1) = 0$.

Next, let us consider any rational numbers

$$p_i = \frac{g_i}{g} \qquad (i = 1, 2, \text{L}, W), \tag{27}$$

where $g_i$ ($i = 1, 2, \text{L}, W$) are any positive integers and $g = \sum_{i=1}^{W} g_i$. The system $A$ is assumed to be described by the probability distribution $\{p_i = g_i/g\}_{i=1, 2, \text{L}, W}$. We construct the system $B$ dependent on $A$ as follows. $B$ contains $g$ events, which are partitioned into $W$ groups: $B_1$, $B_2$, $\text{L}$, $B_W$. $B_j$ ($1 \leq j \leq W$) has $g_j$ events. Once the $i$th event $A_i$ of the system $A$ was found, i.e., $A = A_i$, then, in the system $B$, $g_i$ events of the group $B_{j=i}$ have the same conditional probability $1/g_i$ and all the events of the other groups $B_{j \neq i}$ have the vanishing probability. $S_q$ of $B$ thus constructed is calculated to be

$$S_q[B|A_i] = S_q\left(\frac{1}{g_i}, \frac{1}{g_i}, \mathrm{L}, \frac{1}{g_i}\right) = L_q(g_i) = \frac{1}{1-q}\left[(g_i)^{\lambda(q)} - 1\right]. \tag{28}$$

Therefore, the nonextensive conditional entropy is given by

$$S_q[B|A] = \langle S_q[B|A_i]\rangle_q^{(A)} = \frac{\sum_{i=1}^{W}(p_i)^q S_q[B|A_i]}{\sum_{i=1}^{W}(p_i)^q}$$

$$= \frac{1}{1-q}\left[\frac{\sum_{i=1}^{W}(p_i)^q (g_i)^{\lambda(q)}}{\sum_{i=1}^{W}(p_i)^q} - 1\right]. \tag{29}$$

On the other hand, the composite system $(A, B)$ consists of the events $A_i B_{j=i}$ ($1 \leq i \leq W$). For a given $i$, the number of possible events $A_i B_{j=i}$ ($1 \leq i \leq W$) is $g_i$, and therefore the total number of events in the composite system is $\sum_{i=1}^{W} g_i = g$. The probability of finding the event $A_i B_{j=i}$ is $p_i \times (1/g_i)$, which is the equiprobability $1/g$. Therefore, $S_q$ of the composite system is

$$S_q[A, B] = L_q(g) = \frac{1}{1-q}\left[g^{\lambda(q)} - 1\right]. \tag{30}$$

Substituting eqs. (29) and (30) into [ ]* and using eq. (27), we have

$$S_q[A] = \frac{1}{1-q} \left[ \frac{\sum_{i=1}^{W} (p_i)^q}{\sum_{i=1}^{W} (p_i)^{q+\lambda(q)}} - 1 \right]. \tag{31}$$

This holds for any rational $p_i$ $(i = 1, 2, \text{L}, W)$, but actually for any probability distribution $\{p_i\}_{i=1,2,\text{L},W}$ due to the assumption of continuity in [ ]*.

A remaining task is to determine $\lambda(q)$. For this purpose, it is sufficient to calculate the nonextensive conditional entropy using the form in eq. (31) and impose [ ]* on it. Consequently, we find

$$\lambda(q) = 1 - q. \tag{32}$$

At this stage, we also see that the Ansatz in eq. (22) is in fact justified. Thus, we see that $S_q$ satisfying [ ]*-[ ¡]* is uniquely equal to the Tsallis entropy in eq. (5). (Q.E.D.)

## 4. Concluding remarks

We have constructed the nonextensive conditional entropy in conformity with the Bayes multiplication law and the pseudoadditivity of the Tsallis entropy. We have generalized the Shannon-Khinchin axioms for the Boltzmann-Gibbs entropy to the nonextensive systems. Based on the proposed set of axioms, we have proved the uniqueness theorem for the Tsallis entropy.

Recently, the nonextensive (nonadditive) conditional entropy has been discussed in the quantum context in Ref. [9]. There, it has been shown to give rise to the strongest criterion for separability of the density matrix of a bipartite spin-1/2 system for validity of local realism. We also mention that characterization of the Tsallis entropy has been considered in Ref. [10] from the viewpoint of the concept of "composability", which means that the entropy of the total system composed of statistically independent subsystems is expressed as a certain function of the entropies of such subsystems. (The additivity of the Boltzmann-Shannon entropy and the pseudoadditivity of the Tsallis entropy are the actual examples.) The authors of Ref. [10] has shown that if a quantity satisfies the composability (and some other supplementary conditions), then it is given by the Tsallis entropy with $q > 1$.

## Acknowledgments

The author would like thank Professor A. K. Rajagopal for discussions about diverse topics of nonextensive statistical mechanics. This work was supported in part by the GAKUJUTSU-SHO Program of College of Science and Technology, Nihon University.